# Genetic drift at expanding frontiers promotes gene segregation (Preprint; look up PNAS website for published version)

Oskar Hallatschek[1, 2], Pascal Hersen[2, 3], Sharad Ramanathan[2, 4] and David R. Nelson[1, 2]

[1]*Department of Physics, Harvard University, 02138 Cambridge, MA, USA*

[2]*FAS Center for Systems Biology, Harvard University, 02138 Cambridge, MA, USA*

[3] *CNRS, Laboratoire Matière et Systèmes Complexes, Université Paris VII, Paris, France*

[4] *Bell Laboratories, Alcatel-Lucent, Murray Hill, NJ 07974, USA*

**Competition between random genetic drift and natural selection plays a central role in evolution: Whereas non-beneficial mutations often prevail in small populations by chance, mutations that sweep through large populations typically confer a selective advantage. Here, however, we observe chance effects during range expansions that dramatically alter the gene pool even in large microbial populations. Initially well-mixed populations of two fluorescently labeled strains of *Escherichia coli* develop well-defined, sector-like regions with fractal boundaries in expanding colonies. The formation of these regions is driven by random fluctuations that originate in a thin band of pioneers at the expanding frontier. A comparison of bacterial and yeast colonies (*Saccharomyces cerevisiae*) suggests that this large-scale genetic sectoring is a generic phenomenon that may provide a detectable footprint of past range expansions.**

A principal tenet of modern evolutionary biology is that Darwinian selection and random genetic drift compete in driving evolutionary change (*1*). Numerous investigations (*2-4*) have demonstrated that selection usually prevails over genetic drift in large populations (law of large numbers). A major departure from this behavior occurs, however, when large populations undergo range expansions. The descendents of individuals first settling in a new territory are most likely to dominate the gene pool as the expansion progresses (*5, 6*). Random sampling effects among these pioneers results in genetic drift that can have profound consequences on the diversity of the expanding population. Indeed, spatially varying levels of genetic diversity and colonization patterns appear to be correlated in many species (*7-9*). For example, the often observed south-north gradient in neutral genetic diversity ("southern richness to northern purity" (*10*)) on the northern hemisphere is thought to reflect past range expansions induced by glacial cycles (*9*). While these trends indicate that genetic drift during range expansions has shaped the gene pool of many species, the underlying spatial mechanism remains obscure: Diversity gradients are often difficult to interpret and potentially interfere with the signal of spreading beneficial mutations (*9, 10*). In fact, a major challenge of present-day population genetics is to decide whether natural selection or a past demographic process is responsible for the prevalence of common mutations (*11*).

Here, we use simple microbial systems to study the nature of random genetic drift in range expansions of large populations. We observe chance effects that segregate the gene pool into well-defined, sector-like regions of reduced genetic diversity. The genetic segregation on the population-level is the consequence of number fluctuations on a much smaller scale, within a thin region of reproducing pioneers at the expanding frontier. We expect these patterns to be a general signature of continuous range expansions in populations exhibiting moderate rates of turnover and migration. These results offer a novel means to identify and interpret historically neutral mutations that swept through a population merely by chance.

We used two strains of fluorescently labeled bacteria (*E. coli*) to track both the neutral gene dynamics and the population growth during a range expansion. The strains were genetically identical except for expressing either CFP or YFP, which differ only by a single point mutation, from the same constitutively active promoter. We mixed both strains to obtain cultures with different proportions and placed a droplet containing $\approx 10^6$ cells at the center of an agar plate containing rich growth medium (LB) (*12*). Population growth and mutant distribution were monitored during four days by means of a stereo-microscope. Figure 1 summarizes the colonial development of a founder population of CFP- to YFP-tagged strains at mixing ratios of 50:50 and 95:5. The time series of fluorescence images in Fig. 1A shows that, in our experiments with immotile *E. coli* on hard agar, there is no noticeable temporal change in the mutant distribution behind the population front. The mutant distribution evolves only at the leading edge, much like a (dichromatic) carpet that is knitted from the inside to the outside. The final state of this "genetic" carpet (Fig. 1B and 1C) thus can be taken as a frozen record of the colonization process.

The observed mutant distributions can be partitioned into areas of two very different patterns. The central region of Fig. 1B, which we refer to as the homeland, exhibits a dense speckle pattern reminiscent of the homogeneous and well-mixed initial population. This homeland is bounded by a ring of higher speckle density, which was formed at the edge of the initial droplet (innermost black line in Fig. 1B) during evaporation of its water content (most likely due to flows generated by the evaporation process (*13*), see Fig. S1). From this ring towards the boundary of the mature colony, the population segregates into single-colored domains. A large number of differently labeled flares radiate and gradually coalesce into a few large sectors that reach the edge of the mature colony. Fig. 1C, which corresponds to the unbalanced mixture 95:5 of YFP to CFP, also shows a subdivision into an inner and outer region. The main difference is that the spots and flares are so dilute that their evolution is effectively "non-interacting" and can be individually identified. In particular, only two spots of the minority population evolve into sectors that actually reach the edge of the mature colony.

The formation of morphological sectors, often visible to the naked eye, is a well-known phenomenon in microbial colonies harboring cells in different epigenetic states or carrying non-neutral mutations that occur during growth (*14, 15*). In cases where the incipient microbial population consists of very few individuals (representing a bottleneck (*16*)), sector angles are frequently used to infer the initial composition (e.g., in a colony sectoring assay (*17-19*)).

In contrast, here, sectoring neither arises due to a small founder population (the initial droplet of Fig. 1A roughly contains $10^6$ cells) nor due to mutations and/or epigenetic changes because the genetically segregated domains are non-morphological, *i.e.* indiscernible under white light. Furthermore, growth rate measurements (Fig. S2) show that populations derived from sectors had no significant fitness advantage or disadvantage against the ancestral strain. We argue, instead, that Fig. 1 is a visible manifestation of random genetic drift acting at the leading edge of a range expansion. Genetic drift refers to the fluctuations of genotype frequencies due to the randomness in the reproduction process, as opposed to persistent differences in their phenotype. These fluctuations can ultimately lead to the fixation of neutral, or even deleterious, genes if they are strong compared to mutations and selection. In a permanently well-mixed population grown in a well-shaken test tube, genetic drift is usually weak, because the large number of individuals guarantees an almost deterministic growth of the population (law of large numbers). In the present case of a continual range expansion, however, chance effects are strongly enhanced because the effective population of reproducing individuals is a small fraction of the total population. Only a very thin layer ($\sim 70 \mu$m as inferred from the product of measured growth velocity of $1.3 \times 10^{-2} \mu$m/s, and doubling time of $90$ min, see Fig. S2) of growing cells (the "pioneers") at the boundary of the colony is able to pass on their genes to the next layer of outwardly growing cells. Apparently, the reduction in effective population size due to this continual bottleneck is so strong that the population quickly segregates at the wave front into different domains fixed for one of the two mutant strains.

After the wave front is segregated into monochromatic flares very early in our experiments, a coarsening process occurs: As the colony grows further, some of the flares are squeezed out of the colonization front and others dominate. Very few domains "survive" and reach the edge of the mature colony. From the view point of population genetics, the significance of this coarsening process is that it leads to a gradual decrease in the genetic diversity at the wave front and generates a characteristic sectoring pattern in the neutral genetic makeup of a population. Understanding this process is also of practical importance, as it is a major source of randomness in the relation between the final sector angles and the composition of the incipient microbial population (*19*).

The coarsening dynamics is intimately related with the rugged path of the boundaries separating neighboring domains. A single colored domain looses contact with the wave front when the leading ends of the flanking boundaries meet and annihilate each other (for an illustration, see Fig. S3). Such an unsuccessful sector is, henceforth, trapped in the bulk of the colony without any chance to participate in the further colonization of the agar plate. The sectors surviving this annihilation process, on the other hand, grow in size at the cost of those left behind, and appear to surf (*5, 6*) on the wave front. The irregular shapes of domain boundaries drive the decimation of sectors at the colonization front: Domain boundaries would not collide (and annihilate) if they were straight radial lines. In order to dissect the coarsening mechanism, we study the wandering of individual domain boundaries quantitatively. Suppose, as a null model, that the leading tip of these lines carry out diffusive random walks around their average directions (*20*). Then, the mean square transverse displacement from a straight line should increase linearly with the length $L$ of the random walk. To test this hypothesis, we have plotted the mean square displacement from the best linear fit, extracted from a set of 92 domain walls, as a function of length in Fig. 2B on a double logarithmic scale. Our data indicate that the mean square displacement is indeed well described by a power law, over more than two decades. However, its increase with length $L$ is stronger than linear, i.e., significantly faster than expected for a diffusive random walk; we find the scaling $\sim L^{2\zeta}$ with a wandering exponent $\zeta = 0.65 \pm 0.05 > 1/2$.

The super-diffusive character ($\zeta > 1/2$) of the domain wall motion can be explained by the observation that the shape of the colonial edge locally biases the course of the domain boundary. Indeed, in our experiments we find that despite vigorous zigzag motion, there is a clear correlation between the course of the domain wall and the shape of the colony. Notice from the colored mutant distribution in Figs. 1A and 1B, in which a series of black lines documents the advance of the population front, that the average orientation of domain boundaries seems to be locally perpendicular to the colonial edge. When this locally preferred direction does not coincide with the overall (*i.e.* time-averaged) growth direction, the leading tip of the domain wall will drift transverse to the growth direction, as illustrated in Fig. 3. This drift fluctuates in time, just as the shape of the colonial front does, and has zero mean. By providing a fluctuating bias, a time-dependent interface roughness is thus expected to amplify the random wandering of the domain boundaries. The following scaling argument suggests that this effect can even change the character of the random walk: Suppose the interface kinetics of the colony is characterized (*21*) by a roughness exponent $\chi$, and a dynamical exponent $z$; *i.e.*, an interface segment of linear dimension $L$ exhibits displacements of the order $L^{\chi}$ from a straight line and relaxes on a time scale $L^{z}$. We then argue that, during the time $t$, a transverse drift $\sim L(t)^{\chi-1} \sim t^{(\chi-1)/z}$ acts on the leading tip of a domain wall that is proportional to the tilting angle of a typical undulation on the scale L. This drift entails transverse displacements of the random walk scaling like $t^{1+(\chi-1)/z}$, suggesting a wandering exponent of $\zeta = 1 + (\chi-1)/z$. Our deterministic analysis is valid only in the case of super-diffusion, $\zeta > 1/2$, such that any intrinsic regular diffusion is negligible on sufficiently large times. A simple bacterial growth model, the Eden model[11], is known to be characterized by $\chi = 1/2$ and $z = 3/2$ in one dimension, leading to a wandering exponent of $\zeta = 2/3$ (verified by simulations (*22*)), which is close to the exponent $\zeta = 0.65 \pm 0.05$ found in our experiments (Fig. 2B).

Since the endpoints of domain walls carry out random walks (with the radial axis representing time) and annihilate when they collide, their gradual decimation may be described by a one-dimensional model of annihilating random walkers (*23*) (with periodic boundary conditions). These random walkers, which come in pairs, will annihilate completely for large enough times, if their available space is finite. This scenario applies to a colony which advances approximately as a straight line, as can be induced experimentally through a line-inoculation (cf. Fig. 4C, D). For describing the case of circular microbial colony, such as Fig.1, this analysis has to be modified to account for the fact that the circumference grows linearly in time. Due to this inflation, two domain walls tend to drift apart at a rate proportional to the velocity of the expanding population front. Incorporating this feature into an annihilating walker model with inflation reveals a crossover time beyond which the increase in the size of the circumference is faster then the random walk of the domain walls. As a consequence, domain walls stop colliding and their mutual annihilation ends. Thus, domain walls have a non-zero survival probability in circular colonies. A simple scaling argument balancing the opposing effects of inflation and domain wall wandering indicates that the expected genetic diversity, as reflected by the sectoring pattern, is reminiscent of the shape of the pre-colonial habitat, rather than the initial number of founder cells. In particular, we expect the relationship $N_s \sim r_0^{1-\varsigma}$ between the average number $N_s$ of surviving sectors and the radius $r_0$ of the homeland (supporting online text).

Our neutral model for the coarsening dynamics in expanding *E. coli* colonies suggests that the sectoring phenomenon is not unique to the considered microbial species, but caused by the nature of chance effects during colonization. Indeed, our experiments on growing colonies of haploid *S. cerevisiae* give qualitatively similar results (Fig. 4). Again, we observe an initial coarsening phase, in which domain boundaries annihilate, followed by a stationary phase, in which the sectoring pattern is stable. Compared to *E. coli*, however, yeast colonies have a much larger number of surviving sectors and much straighter domain boundaries. Both observations are consistent with our model of annihilating domain boundaries, because the frequency of annihilation events is expected to decrease for straighter domain boundaries. The disparate domain wall wandering could be due to differences in the mode of replication, the shape of the cells or the thickness of the colonies. However, even though the local genetic patterns in colonies of different microbial species might be quite different, the genetic segregation on larger scales seems to be a more general consequence of a microbial range expansion.

In this report, we have shown that the genetic makeup of large microbial populations can be dramatically changed by enhanced genetic drift at expanding population fronts. Neutral mutants quickly segregate into monoclonal domains that further coarsen as the colony grows. The final sectoring patterns observed in circular colonies are controlled by a balance between deterministic inflation of the colonial perimeter, dominant at large times, and the stochastic meandering of domain boundaries causing them to annihilate on short times.

In light of inferring past migrations from spatially resolved genetic data, our results suggests that, apart from a much anticipated general reduction of genetic diversity by a range expansion, a fragmentation of the colonized regions into sectors has to be considered, which is stabilized during expansion by an inflationary effect. Although, demonstrated here only for non-motile microbial species, we believe that the underlying segregation mechanism be quite general for populations growing continuously and isotropically in two dimensions (*e.g.*, a viral epidemic spreading through its host population (*24*)). As long these genetic patterns are not blurred by subsequent dispersal and population turnover, they provide a record of the colonization dynamics as well as information about the pre-colonial habitat.

**Acknowledgements** It is a pleasure to acknowledge T. Shimizu for helpful discussions and, together with H. Berg and V. Sourjik, for providing the *E. coli* strains and plasmids; the O'Shea lab for the *S. cerevisiae* strains; M. Bathe, K. Foster, M. Nowak, B. Stern, J. Wakeley and J. Xavier for providing comments on the manuscript. This research was supported by the German Research Foundation through grant no. Ha 5163/1 (OH), the National Science Foundation through grant DMR-023163 and the Harvard Materials Research Science and Engineering Center through grant DMR-D2138D5 (DRN), the Human Frontiers Grant (SR), Keck Futures Initiative Grant (SR) and an NIGMS center grant (PH, SR).

**Author Information** Correspondence and requests for materials should be addressed to D.R.N. (e-mail: nelson@cmt.harvard.edu, tel: 617-495-4331, fax: 617-496-2545)

**Supporting online Material**
**www.sciencemag.org**
Materials and methods
Figs. S1, S2, S3.

**Fig. 1.** Fluorescent images of bacterial colonies grown from a mixture of CFP and YFP labeled cells reveal spatial segregation of neutral genetic markers. **(A)** Images taken at twelve hour intervals, of a growing colony founded by a 50:50 mixture of bacteria ($\approx 10^6$ *E. coli* cells) carrying plasmids with either CFP or YFP. Even though the bacteria were otherwise genetically identical, the growing colony shows complete segregation of the two neutral markers (CFP and YFP) over time. The progression of fluorescent images suggests that the dynamics of segregation is restricted to the edges of the colony, while, except for a gradual thickening, the interior distribution of CFP and YFP is essentially frozen. The bright field image at 0h indicates the initial extent of the founding population, which is mostly lined up on a circle (shown enlarged in Supplementary Figure 1; the central dark spot is a surface irregularity caused by the delivering pipette). Within 12 hours and 72 hours, the YFP signal is shown; at 86 hours (final time point), YFP (green) and CFP (red) images are overlaid on each other with the physical boundary (black lines) of the growing colony at earlier times obtained from bright light images. **(B)** Enlarged view at 84h of the founding region (innermost black circle) and its immediate surroundings. **(C)** Fluorescence image of a colony produced by a founder population ($\approx 10^3$ cells) with 95% YFP and 5% CFP labeled mutants taken 62 hours after inoculation. The magnified region shows that, by chance, a sector of the minority population (CFP) can successfully "surf" (*4, 5*) even in this imbalanced case.

**Fig. 2.** Random meandering of domain boundaries is super-diffusive. **(A)** From fluorescent images of *E. coli* colonies, domain boundaries were extracted (red lines) using an edge detection algorithm. **(B)** To test whether these domain boundaries resemble time traces of diffusive random walks, we quantified their contour fluctuations as a function of contour length $x$. To this end, we chose a sliding window of size $x$, parameterized the contour by the transverse displacements $y$ from the linear best fit, and determined the mean square displacement $\overline{y^2}$ (*12*). **(C)** The mean square transverse displacement is shown as a function of the size $x$ of the sliding window. 92 domain boundaries sampled from 12 different *E. coli* colonies were analyzed (black points). Even though the data for individual domain boundaries (black lines) are quite noisy,

the ensemble average (red crosses) very closely follows a straight line in this double logarithmic plot. The exponent $2\zeta = 1.31 \pm 0.1$ of the fitted power law (blue line) indicates that boundaries carry out super-diffusive random walks around their average growth direction. I.e., they wander more vigorously than conventional random walks, which have a wandering exponent given by $2\zeta = 1$.

**Fig. 3.** The roughness of the colonization front (black line) influences the wandering of locally perpendicular domain boundaries (red lines). In this sketch, the middle domain boundary (dashed blue) tends to follow the blue arrow indicating the local growth direction of the colony. Due to the stochastic surface growth, the local growth direction deviates from the average growth direction (black dashed arrow). Consequently, the domain boundary is subject to a drift (red arrow) transverse to the average growth direction, which is proportional to the local tilt of the interface. This tilt depends upon the roughness of the edge of the colony, which we assume to be characterized by a roughness exponent (*21*) $\chi$: on a length scale $\lambda$, the edge of the colony exhibits stochastic shape undulations of amplitude $h \sim \lambda^{\chi}$. The corresponding tilt, $h / \lambda \sim \lambda^{\chi-1}$, biases the path of the considered domain boundary until the surface irregularity causing it has relaxed. Due to the roughness of the colonial edge, the domain boundaries thus experience a fluctuating bias, which may significantly change their wandering statistics, as explained in the text.

**Fig. 4.** Segregation patterns of two different microbial species are qualitatively similar, but different in detail. **(A)**, **(B)**, Both bacteria (*E. coli*) and yeast (*S.*

*cerevisiae*) colonies exhibit spatial gene segregation when they grow colonies on agar plates. However, the number of (surviving) sectors is much larger in yeast colonies. For the linear inoculations **(C)**, **(D)**, the petri dish was gently touched with a sterile razor blade that was previously wetted by a liquid culture of a binary mixture of mutants. **(E)**, **(G)**, Continuous patches of boundary regions and homeland (bounded by dashed line) at a magnification of 51x for *E. coli* and yeast, respectively. **(F)**, **(H)**, Tip of a sector that dies out (*E. coli*) and section boundary at the frontier (yeast), respectively, with single cell resolution (100x).

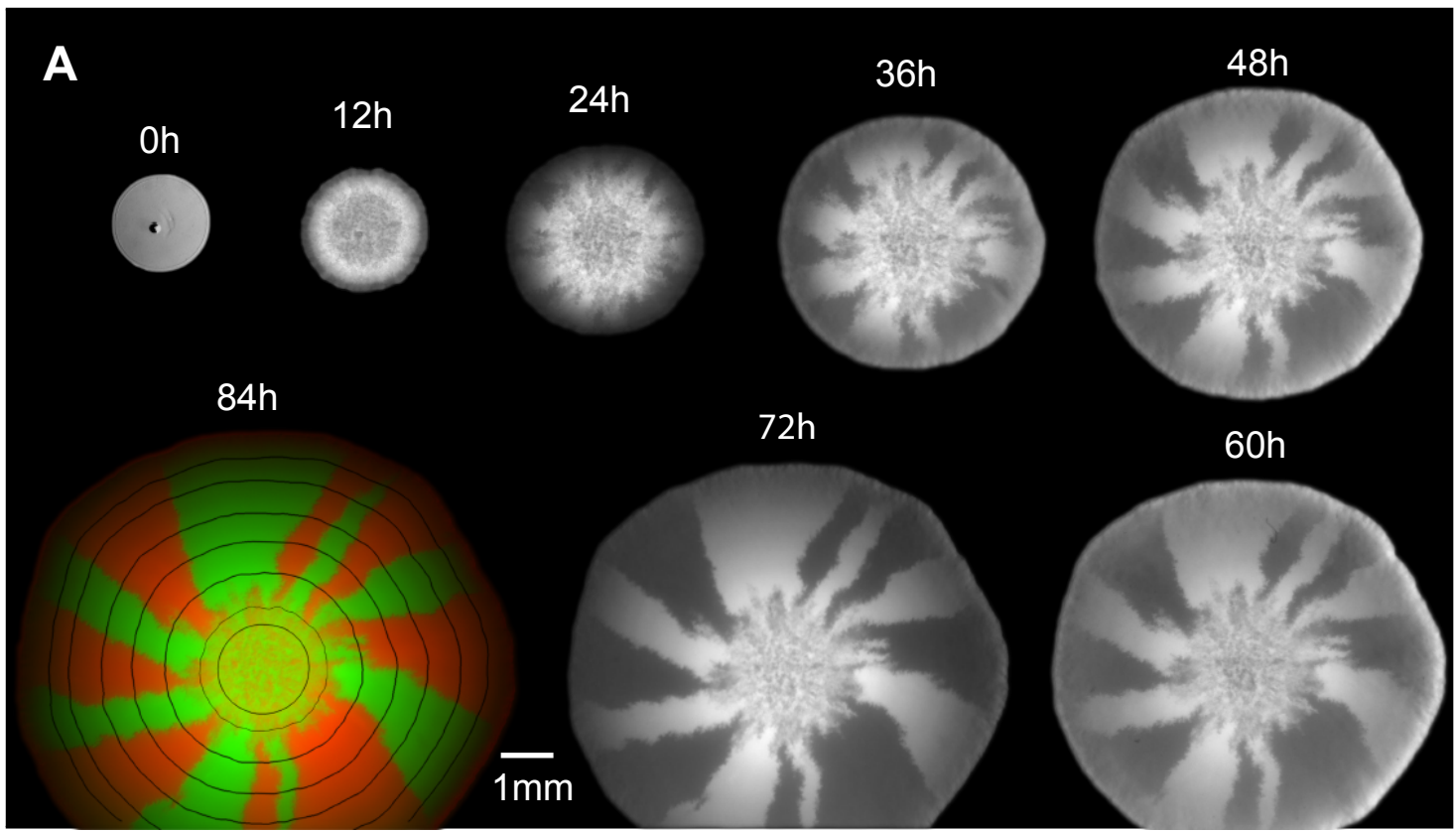

A

0h  12h  24h  36h  48h

84h  72h  60h

1mm

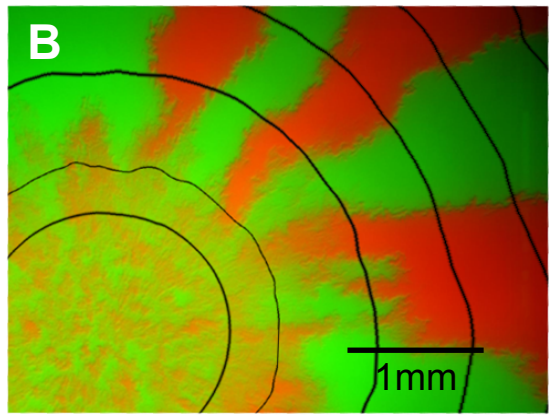

B

1mm

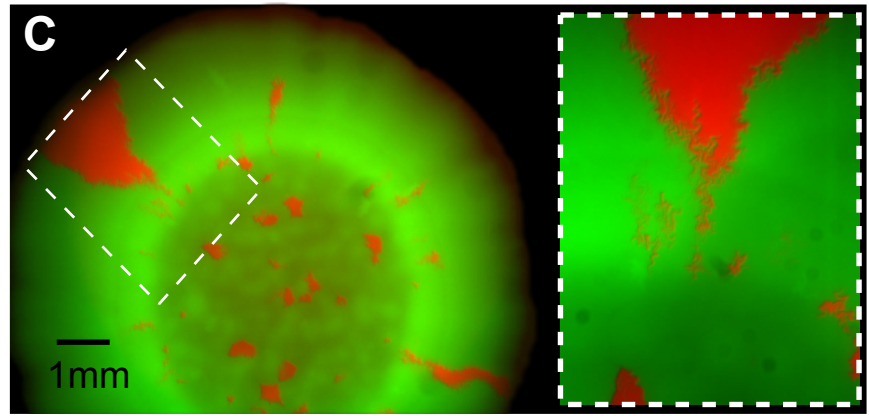

C

1mm

Figure 1

O. Hallatschek

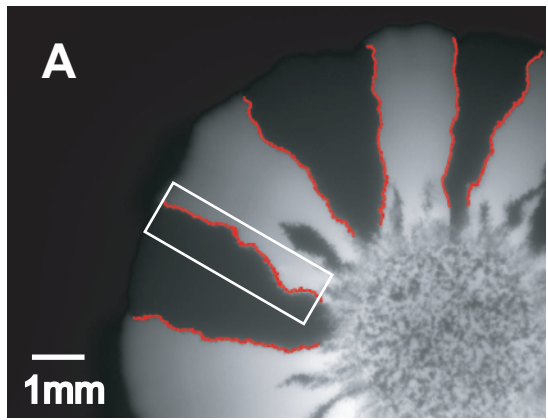

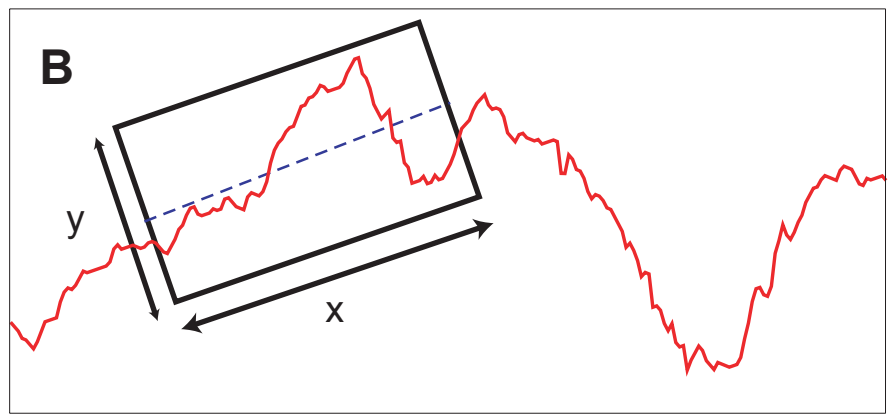

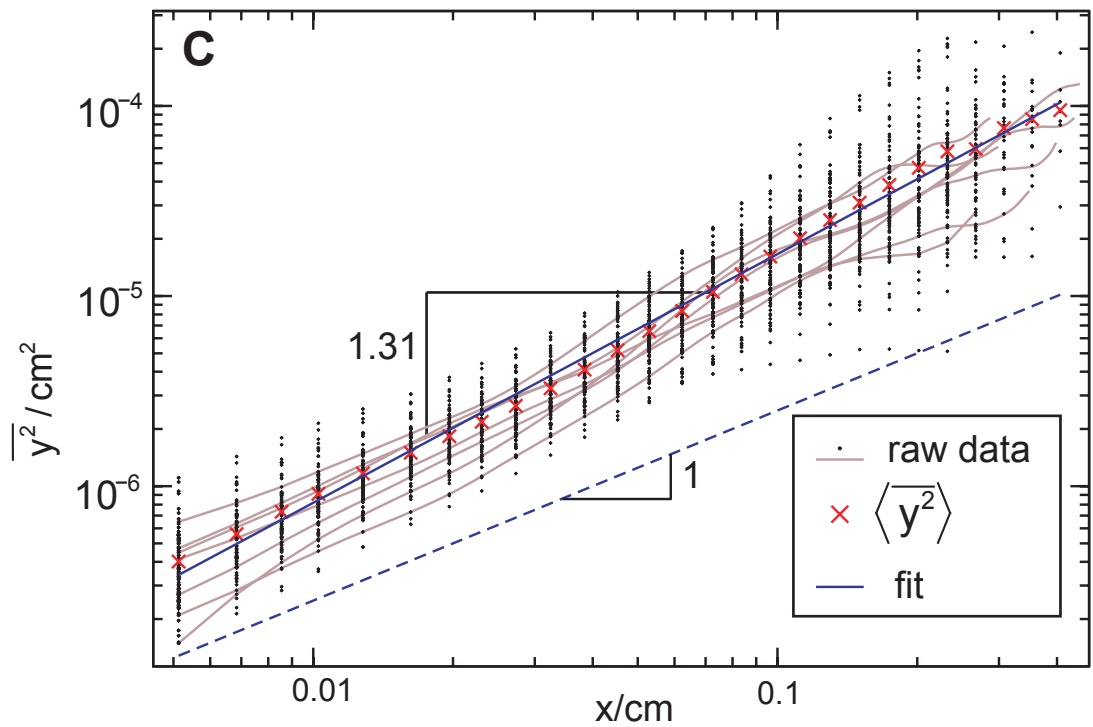

Figure 2

O. Hallatschek

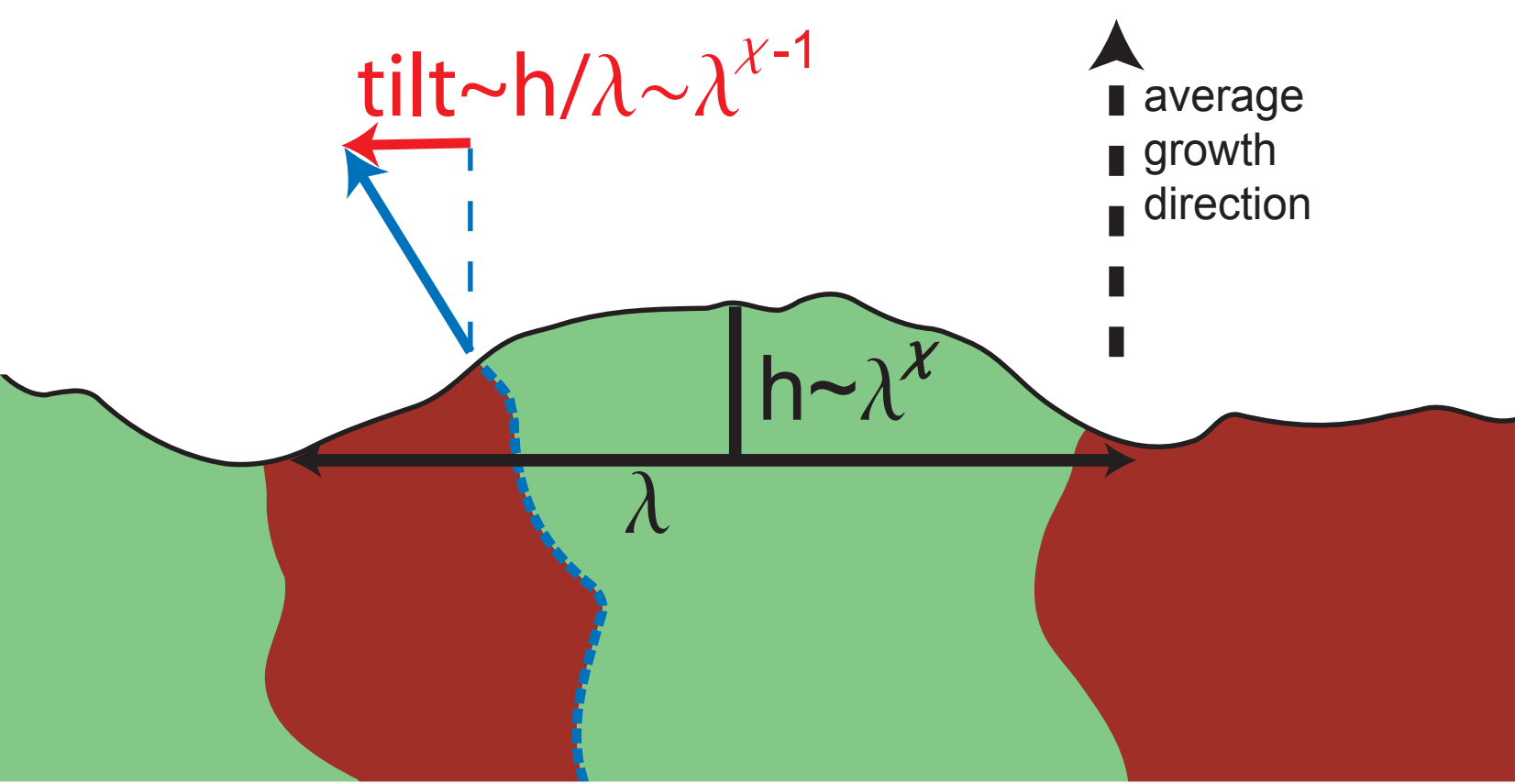

$tilt \sim h/\lambda \sim \lambda^{\chi-1}$

$h \sim \lambda^{\chi}$

$\lambda$

average growth direction

Figure 3                    O. Hallatschek

*E. coli*  |  *S. cerevisiae*

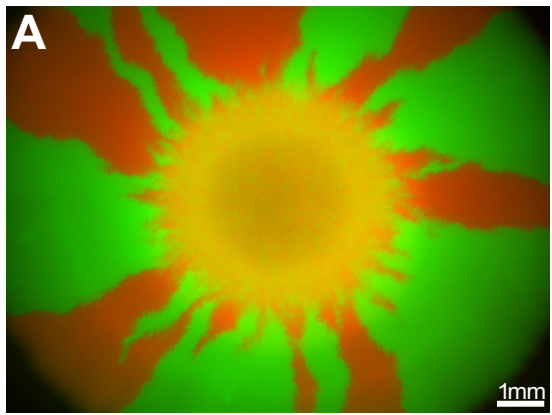
**A**
1mm

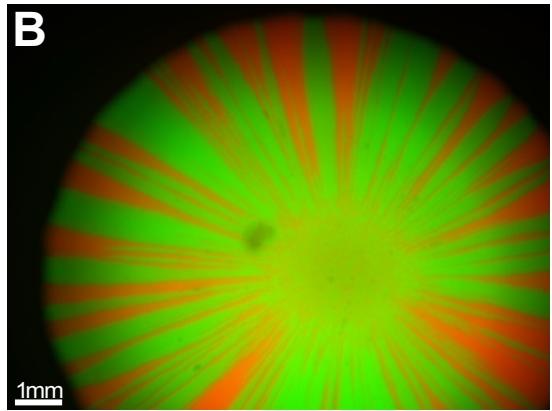
**B**
1mm

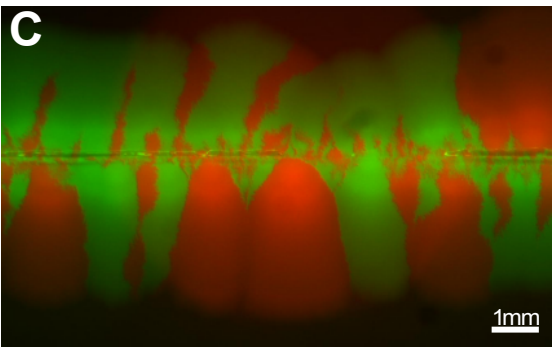
**C**
1mm

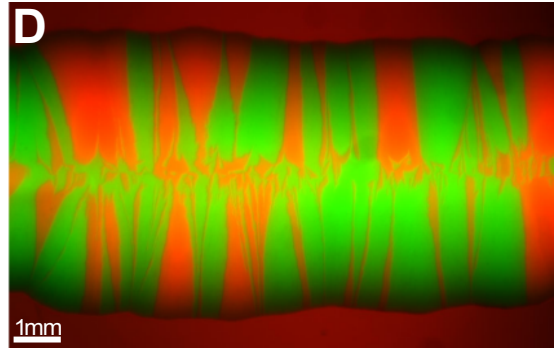
**D**
1mm

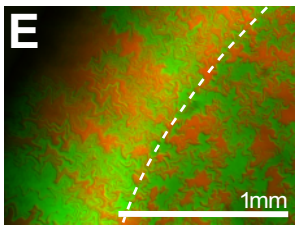
**E**
1mm

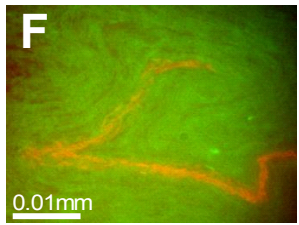
**F**
0.01mm

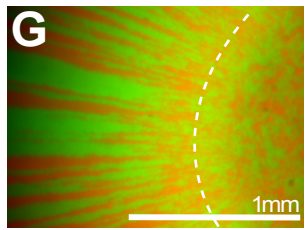
**G**
1mm

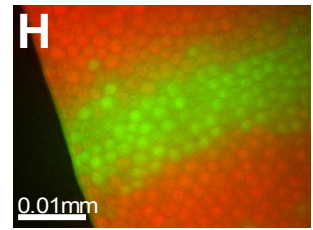
**H**
0.01mm

Figure 4  O. Hallatschek